\documentclass[aps,prd,preprint,groupedaddress,epsfig]{revtex4}
\usepackage{graphicx}
\newcommand{\be}{\begin{equation}}
\newcommand{\ee}{\end{equation}}
\def\rs{\mbox{$\sqrt{s}$}}
\def\pbarp{\mbox{$ \bar{p}p$}}

\newcommand{\ba}{\begin{array}}
\newcommand{\ea}{\end{array}}
\newcommand{\beqa}{\begin{eqnarray}}
\newcommand{\eeqa}{\end{eqnarray}}
\newcommand{\PL}[3]{{ Phys. Lett.}       {\bf #1} {(19#2)} {#3}}
\newcommand{\PRL}[3]{{ Phys. Rev. Lett.} {\bf #1} {(19#2)} {#3}}
\newcommand{\PR}[3]{{ Phys. Rev.}        {\bf #1} {(19#2)} {#3}}

\newcommand{\NP}[3]{{ Nucl. Phys.}       {\bf #1} {(19#2)} {#3}}

\begin{document}
%\begin{titlepage}
\preprint{IISc-CHEP/9/04}
\preprint{hep-ph/0408355}
\title{Soft gluon radiation  and energy dependence of 
 total hadronic cross-sections}
\author{  
R.M. Godbole}
\email{rohini@cts.iisc.ernet.in}
\affiliation{ Center for High Energy Physics, Indian Institute of Science, 
Bangalore,  560 012, India}
\author{A. Grau}
\email{igrau@ugr.es}
\affiliation{Departamento de Fisica Teorica y del Cosmos, Universidad de 
Granada, Spain}
\author{G. Pancheri}
\email{pancheri@lnf.infn.it}
\affiliation{INFN, Laboratori Nazionali di Frascati, P.O. Box 13,
I-00044 Frascati, Italy}
\author{Y. N. Srivastava}
\email{srivastava@pg.infn.it}
\affiliation{Physics Department and INFN, University of Perugia, 
Perugia, Italy}
%**********************
\date{\today}
\begin{abstract}
        An impact parameter representation for soft gluon radiation
is applied to obtain both the initial decrease of the total cross-section 
($\sigma_{tot}$) for proton-proton collisions as well as the later 
rise of $\sigma_{tot}$ with energy for both  $pp$ and $p\bar{p}$. The 
non-perturbative soft part of the eikonal includes only limited low 
energy gluon emission and leads to the initial decrease in the proton-proton 
 cross-
section. On the other hand, the rapid rise in the hard, perturbative 
jet part of the eikonal is tamed into the experimentally observed 
mild increase by soft gluon radiation whose maximum energy 
rises slowly with energy.
\end{abstract}
\maketitle
\section{Introduction}
In this paper, we extend our approach to  the role played by soft gluon 
radiation in determining the energy dependence of total cross-sections
\cite{ff2}, by including  new low energy features. The  
 experimental information on the total cross-section   is
now  sufficient to allow for definite progress  towards 
its description through QCD. Thanks to the recent measurements at 
HERA~\cite{HERAZ,HERAH1,DIS,camil} and LEP~\cite{L3,OPAL} providing
knowledge of total hadronic cross-sections involving photons  in the energy 
interval
$\sqrt{s}=1\div 100\ GeV$, we now possess a complete set of processes to
study in detail and in depth, namely $p p$, 
$p{\bar p}$, $\gamma p$ and $\gamma \gamma$. The purely 
hadronic processes are well measured 
over an extended range, up to cosmic 
ray energies \cite{PDG05}, leading to quite precise 
parametrizations \cite{martincosmic},
while the other two allow for probes of the hadronic content of the
photon versus that of the proton. They also allow for checks of the
 Gribov-Froissart factorization hypothesis \cite{gribov,martink}.

  The three basic features which the data exhibit and which require a 
theoretical explanation and understanding are: 
(i) the normalization of the cross section, (ii) an initial decrease and 
(iii) a  subsequent rise with energy. %\cite{therise}. 
All three are reasonably well described in the Regge trajectory language, 
which 
%have suggested 
suggests a parametrization of all total cross-sections\cite{DL} 
as a sum of powers of the square of the c.m. energy
$s$. The oldest and simplest of these parametrizations is   
\begin{equation}
\sigma_{tot}(s)=Xs^\epsilon+Ys^{-\eta}.
\label{DL}
\end{equation}
In this model, the initial decrease reflects the disappearance 
(with increasing energy) of a Regge trajectory exchanged in the t-channel, 
with $\eta=1-\alpha_\rho(0)$ where the intercept (at $t\ =\ 0$) of the 
leading Regge trajectory is $\alpha_\rho(0)\approx 0.5$.  At the same time, 
the rise in the cross section is attributed to the exchange of a trajectory 
(the Pomeron) with the quantum numbers of the vacuum, such 
that $\alpha_P(0)=1+\epsilon$.  $\epsilon$ is expected to be a
small number so as not to defy too much the
Froissart bound, and, phenomenologically, 
$\epsilon= 0.08\div 0.12$\cite{PDG,jhep}.

The two terms in which the cross-sections 
are split in Eq.(\ref{DL}) describe  well 
%all 
the main features of
the observed total cross-section data, with an apparently
constant behaviour at $\sqrt{s}\approx
{{\eta}\over{\epsilon}}{{Y}\over{X}}$.  For proton-proton and
 proton-antiproton scattering, this occurs between 15 and 25 GeV, 
where the cross-section is about 40 mb. 
At large energies, the Regge term disappears, while the first term 
is important everywhere, and
 dominates asymptotically. Thus the decrease is due   to the 
Regge term,
the rise described by the Pomeron with some taming due to the Regge term 
and the normalisation is determined by both the Regge and the Pomeron terms.
% the rise 
% to the Pomeron, with some taming due to 
%the Regge term, and  the normalization is due 
%to both the Regge and  Pomeron terms. 
More recent analyses add further terms  such as $log^2(s/s_0)$
terms\cite{newfit}.

%       In the following we shall discuss these features one by 
%one within the context of QCD, using Eq.(\ref{DL}) as 
%reference. This parametrization is  a useful toy model with well
%grounded theoretical justifications in general features of scattering
%amplitudes, like  analyticity and crossing. In the present work, we develop 
%further the  model of ref.\cite{ff2} to increase our understanding of the 
%energy dependence within QCD. 
% REPLACRD BY
In the following we shall discuss these features one by 
one within the context of QCD, using the simple but useful parametrization
of Eq.(\ref{DL}) as reference. We shall modify it and through it 
develop further the  model of ref.\cite{ff2} to increase our understanding 
of the energy dependence within QCD.   
%{\it
 In  section II, we discuss the QCD origin of the two component model of 
Eq. (\ref{DL}).
In section III, we summarise how the 
%QCD contribution 
Born term of the QCD contribution,  the mini-jets, 
is embedded in the eikonal formalism which 
provides a unitarised description 
%of the rise with energy 
of the total cross-section. In section IV, we discuss the analyticity 
requirements upon the impact parameter amplitudes of the eikonal formalism. 
%{\it 
In section V, we review the soft gluon transverse momentum
  distribution on which our model 
for the b-distribution of partons is based.
%} 
In section VI we discuss the main features of  the  model\cite{ff2} for 
the energy dependence of the impact parameter distribution, induced by 
soft gluon emission. In section VII, through soft gluon emission, we 
obtain the transverse parton overlap function for the non-perturbative 
part of the cross-section. In section VIII,  we introduce
%our choice of the last non-perturbative parameters of the model, viz., 
the normalisation of the cross-section. We  present
our model  for $\sigma_{tot}$ for $pp$ and $p\bar p$ 
and compare them with currently available data. In section IX,
we present an interesting feature of our model: in a proton or antiproton
collision, the (average) distance between the scattering centres (i.e. the 
constituents) in the transverse space decreases as the energy increases.  
Similar results have been previously obtained in an analysis of the hadronic 
events at the Tevatron, through an impact parameter picture.
We note again that our model of the energy dependent impact parameter 
distribution offers a  
reasonable understanding of  both the initial 
fall and later rise with energy of the $\sigma_{tot}$, whereas for 
the normalization there is still further work to do.

 %{\it 
 Before beginning the detailed discussion, we indicate below briefly and 
 qualitatively, why we believe that it is both important and necessary to 
 include effects of soft-gluon emission in the mini-jet 
 formalism \cite{therise,CR,halzen1}. We recall 
 here that in the mini-jet picture the rise in cross-sections is driven by 
the increasing number of low-x gluon-gluon collisions and that the predicted 
rate of the rise is generally found to be uncomfortably large. In section VI
 we give the details of the mechanism by which the effect of soft gluon 
 emissions can reduce the rate of this rise, at any given c.m. energy. Below
we summarise the effect qualitatively.

 In a calculation in the QCD improved 
parton model, the effect of gluon radiation on the longitudinal momentum 
carried by the partons is included, at least in part, in the 
Dokshitzer, Gribov, Lipatov, Altarelli and  Parisi (DGLAP) \cite{DGLAP} evolved 
parton distribution. Here we crucially look at the effect of
initial state 
 gluon radiation on the transverse momentum distribution of partons. In 
our model the soft gluon resummed transverse momentum distribution of  
partons in the hadrons and the parton distribution in  impact parameter 
space are Fourier Transforms of each other.
%} {\bf 
The larger the transverse momentum 
the larger is the acollinearity of the two colliding partons, leading to a 
reduction in parton luminosity and hence to a reduction of the
cross-section.
%}{\it 
The higher the c.m. energy of the parent hadron, the more energetic are the 
parent valence quarks emitting gluons and the more is the acollinearity of the
two partons involved in the parton subprocess. Hence, we  expect the effect
to be dependent on the c.m. energy.
%}

%{\it 
\section{The two component model }

        Before entering into the technical details about unitarization,
it is well to ask (i) where the  ``two component'' structure 
of Eq.(\ref{DL}) comes
from and (ii) why the difference in the two powers (in $s$) is 
approximately a half.

 Let us first remember that the two terms of Eq.(\ref{DL}) reflect the
  well known duality between resonance and Regge pole exchange on the one hand
  and background and Pomeron exchange on the other, established in the late
  60's
  through FESR \cite{FESR}. This correspondance  meant that, while at 
  low energy the cross-section could be written as due to a  background term
  and a sum of resonances,   at higher energy it could be written as 
a sum of Regge trajectory exchanges and   a Pomeron exchange.  

Our present knowledge of QCD description of
  hadronic phenomena  gives    further insight into the nature of these two
  terms.
%  and we  can then use quark hadron
%  duality, to see how at low and intermediate energies the properties of 
%quarks  and gluons manifest themselves as resonances and background 
%or as Regge
%  and Pomeron exchange.}
We shall start answering the above two questions through  considerations 
about the 
bound state nature of hadrons which necessarily transcends perturbative 
QCD. For hadrons made of light quarks($q$) and glue($g$), the two terms 
arise from $q\bar{q}$ and $gg$ excitations. For these, the energy is given 
by a sum of three terms: (i) the rotational energy, (ii) the Coulomb 
energy and (iii) the ``confining'' energy. If we accept the Wilson area 
conjecture in QCD, (iii) reduces to the linear potential\cite{Widom1,
Widom2}. 
Explicitly, in the CM frame of two massless particles,
either a $q\bar{q}$ or a $gg$ pair separated by a relative 
distance r with relative angular momentum $J$, the energy is given by
\be \label{string1}
E_i(J, r) = {{2J}\over{r}} - {{C_i \bar{\alpha}}\over{r}} + C_i \tau r,
\ee     
where $i\ =\ 1$ refers to $q\bar{q}$,  $i\ =\ 2$ refers to $gg$, $\tau$
is the ``string tension'' and the Casimir's are $C_1\ =\ C_F\ =\ 4/3$, $C_2\
=\ C_G\ =\ 3$. $\bar{\alpha}$ is the QCD coupling constant evaluated at a
some average value of r and whose  value
will disappear in the ratio to be considered. The hadronic
rest mass for a state of angular momentum $J$ is then computed through
minimising the above energy
\be \label{string2}
M_i(J) = Min_r[ {{2J}\over{r}} - {{C_i \bar{\alpha}}\over{r}} + C_i \tau r ],
\ee 
which gives
\be \label{string3}
M_i(J) = 2 \sqrt{(C_i \tau)[2J - C_i \bar{\alpha}]}.
\ee
The result may then be inverted to obtain the two sets of linear Regge 
trajectories $\alpha_i(s)$ 
%(which is {\it not} the coupling constant)
\be \label{string4}
\alpha_i(s) = {{C_i \bar{\alpha}}\over{2}} + ({{1}\over{8 C_i \tau}}) s
= \alpha_i(0) + \alpha_i' s.
\ee 
Thus, the ratio of the intercepts is given by
\be \label{string5}
{{\alpha_{gg}(0)}\over{\alpha_{q\bar{q}}(0)}} = C_G/C_F = {{9}\over{4}}. 
\ee
Using our present understanding that resonances are $q{\bar q}$ bound 
states while the background, dual to the Pomeron,
 is provided by gluon-gluon exchanges\cite{landshoff},  the above 
equation can be rewritten  as 
\be \label{string5qhd}
{{\alpha_{P}(0)}\over{\alpha_R(0)}} = C_G/C_F = {{9}\over{4}}. 
\ee
If we restrict our attention to the leading Regge trajectory, namely
the degenerate $\rho-\omega-\phi$ trajectory, 
then  $\alpha_R(0)=\eta\ \approx\ 0.48-0.5$, and we obtain
for $\epsilon\ \approx\ 0.08-0.12$, a rather  satisfactory value. 
The same argument for the slopes gives 
\be \label{string6}
{{\alpha_{gg}'}\over{\alpha_{q\bar{q}}'}} = C_F/C_G = {{4}\over{9}}.
\ee 
so that if we take for the Regge slope $\alpha_R'\ \approx\ 0.88-0.90$, we get
for $\alpha_P'\ \approx\ 0.39-0.40$,  in fair
agreement with lattice estimates \cite{lattice}.

%We are not aware of any alternative explanation for these facts: neither 
%for the need of two components nor for the ratio of the two intercepts.   

We now have good reasons for a break up of the amplitude into 
two components. To proceed further, it is necessary to realize that 
precisely because massless hadrons do not exist,  Eq.(\ref{DL}) violates 
the Froissart bound and thus must be unitarized. To begin this task, 
let us first rewrite  Eq.(\ref{DL}) by putting in the  ``correct'' dimensions
\be \label{DL1} 
\bar{\sigma}_{tot}(s)= \sigma_1 (s/\bar{s})^\epsilon+ \sigma_2
(\bar{s}/s)^{1/2},
\ee
where we have imposed the nominal value $\eta\ =\ 1/2$. In the following,
we shall obtain rough estimates for the size of the parameters in 
Eq.(\ref{DL1}). 

A minimum occurs
in $\bar{\sigma}_{tot}(s)$ at $s\ =\ \bar{s}$, for $\sigma_2\ =\ 2\epsilon
\sigma_1$.
%, so that
If we make this choice, then Eq.(\ref{DL1}) has one less parameter 
and it reduces to
\be \label{DL2} 
\bar{\sigma}_{tot}(s)= \sigma_1 [(s/\bar{s})^\epsilon+ 2 \epsilon
(\bar{s}/s)^{1/2}].
\ee
We can isolate the rising part of the cross-section by rewriting the above
as
\be \label{DL3}
\bar{\sigma}_{tot}(s)= \sigma_1 [ 1 + 2\epsilon(\bar{s}/s)^{1/2}] 
+ \sigma_1 [(s/\bar{s})^\epsilon - 1].
\ee
Eq.(\ref{DL3}) separates cleanly the cross-section into two parts: the first
part is a ``soft'' piece which shows a saturation
to a constant value (but which contains no rise) and the second a ``hard'' 
piece which has all the rise. Morover, $\bar{s}$ naturally provides the 
scale beyond which the cross-sections would begin to rise. Thus, our
``Born''  term assumes the generic form
\be \label{DL4}
\sigma_{tot}^B(s)= \sigma_{soft}(s) + \vartheta (s - \bar{s})
\sigma_{hard}(s).
\ee  
with $\sigma_{soft}$ containing a constant ( the ``old'' Pomeron
with $\alpha_P(0)\ =\ 1$) plus a (Regge) term decreasing
as $1/\sqrt{s}$ and with an estimate for their relative magnitudes
($\sigma_2/\sigma_1\ \sim\ 2\epsilon$). We shall assume that the rising 
part of the cross-section $\sigma_{hard}$ is provided by jets which are 
calculable by perturbative QCD, obviating (atleast in principle) the 
need of an arbitrary parameter $\epsilon$.

        An estimate of $\sigma_1$ may also be obtained through the hadronic
        string 
picture. Eq.(\ref{string2}) gives us the mean distance between quarks
or the ``size'' of a hadronic excitation of angular momentum $J$ in 
terms of the string tension
\be \label{string7}
\bar{r}(J)^2 = {{2J - C_F\bar{\alpha}}\over{\tau}}.
\ee
So the size $R_1$ of the lowest hadron (which in this Regge string
picture has $J\ =\ 1$, since $\alpha_R(0)\ =\ 1/2$) is given by
\be \label{string8}
R_1^2 = {{1}\over{\tau}} = 8 \alpha'
\ee 
If two hadrons each of size $R_1$ collide, their effective radius
for scattering would be given by
\be \label{string9}
R_{eff} = \sqrt{R_1^2 + R_1^2} = \sqrt{2} R_1,
\ee     
and the constant cross-section may be estimated (semi-classically) to be roughly
\be \label{string10}
\sigma_1 = 2 \pi R_{eff}^2 = 4 \pi R_1^2 \approx {{4\pi}\over{\tau}}
= 32 \pi \alpha', 
\ee
which is about $40\ mb$, a reasonable value. 
In the later sections, for the ``soft'' cross-section we shall take a value
        of this order of magnitude  as the nominal value.

        The unitarization now proceeds very simply by eikonal
        exponentiation in
impact parameter space, as described in detail in the sections to follow.
%below). This will require 
%a model for the the impact paameer distribution of partons.  with
%different impact parameter distributions for the two pieces reflecting
%a different matter distribution for each. 
%In this context, it must be 
%borne in mind that for hadronic physics, parameters which ensue reflect 
%the lowest hadronic mass exchanges and thus may be rather removed from the 
%QCD scale.

%\end{document}

\section{The rise with energy}
QCD offers 
%a very 
an elegant explanation of the rise in the minijet formalism as has been 
pointed out by several authors in the past \cite{therise,CR,halzen1}. 
The 
%suggestion 
original suggestion was that the rise of $\sigma_{tot} $ with energy is
driven by the rapid rise with energy of the inclusive jet cross-section
\be \label{sigjet}
\sigma^{ab}_{\rm jet} (s) = \int_{p_{tmin}}^{\rs/2} d p_t
\int_{4 p_t^2/s}^1 d x_1 \int_{4 p_t^2/(x_1 s)}^1 d x_2 \sum_{i,j,k,l}
f_{i|a}(x_1) f_{j|b}(x_2) \frac { d \hat{\sigma}_{ij \rightarrow kl}(\hat{s})}
{d p_t},
\ee
where subscripts $a$ and $b$ denote particles ($\gamma, \ p, \dots$), 
$i, \ j, \ k, \ l$ are partons and $x_1,x_2$ the fractions of the 
parent particle momentum carried by the parton. $\hat{s} = x_1 x_2 s$  and
$\hat{ \sigma}$ are hard partonic scattering cross--sections. Note that
$d \hat{\sigma} / d p_t \propto p_t^{-3}$; the cross--section defined in
Eq.(\ref{sigjet}) therefore depends very sensitively on 
%\ptmin
$p_{tmin}$, which is
supposed to parametrize the transition from perturbative to nonperturbative
QCD.  The rise of the inclusive jet cross-section with energy is understood in 
terms of the increasing number of hard partons which gives rise to an 
increasing probability for the occurrence of hard scattering processes.
Quantitatively, factorisation of QCD allows us to use the currently available 
parametrisations of the scale dependent parton densities and calculate the 
energy dependence of the resulting jet cross-section, by convoluting the
parton densities with the subprocess cross-section determined by perturbative
QCD. 

For any fixed $p_{tmin}$, typically $1\div 2$ GeV, one finds that this 
cross-section is a steeply rising function of energy. 
If $\rs \gg p_{tmin}$, the integral in Eq.(\ref{sigjet}) receives its dominant
contribution from $x_{1,2} \ll 1$. The relevant parton densities can then be
approximated by a simple power law, $f \propto x^{-J}$. In case of $pp$ or
\pbarp\ scattering, $a=b$ and the cross--section asymptotically scales like
\cite{powerlaw}
\be 
\sigma_{\rm jet} \propto \frac {1} {p^2_{tmin}} \left( \frac {s}
{4 p^2_{tmin}} \right)^{J-1} \log \frac {s} {4 p^2_{tmin}},
\ee
if $J>1$. For $J \simeq 1.3$, as measured by HERA, the jet cross--section
will therefore grow much faster than the total \pbarp\ cross--section, which
only grows $\propto \log^2 s$ (Froissart bound \cite{froissart}), or,
phenomenologically \cite{DL} for $\rs \leq 2$ TeV, $\propto s^{0.08}$.
Eventually the jet cross--section (\ref{sigjet}) will therefore 
exceed the total \pbarp\ cross--section. In fact, this rise is far more 
violent than the experimentally observed 
gentle rise of the total cross-section\cite{halzen1,triplegluon}. This
has led to  various phenomenological strategies (among them  the eikonal 
formalism \cite{durand}) directed towards softening  this rise. 
The apparent paradox is solved by the observation that, by definition,
inclusive cross--sections include a multiplicity factor. Since a hard partonic
scattering always produces a pair of (mini--)jets, we can write
\be \label{e5.3}
\sigma^{ab}_{\rm jet} = \langle n_{\rm jet \ pair} \rangle
\sigma^{ab}_{\rm inel},
\ee
where $\langle n_{\rm jet \ pair} \rangle$ is the average number of
(mini--)jet pairs per inelastic collision. $\sigma_{ab}^{\rm jet} >
\sigma^{ab}_{\rm inel}$ then implies 
$\langle n_{\rm jet \ pair} \rangle > 1$, which means that, on average, 
each inelastic event contains more than one hard
partonic scatter. The simplest possible assumption about these multiple
partonic interactions is that they occur completely independently of each
other, in which case $n_{\rm jet \ pair}$ obeys a Poisson distribution. At a
slightly higher level of sophistication, one assumes these interactions to
be independent only at fixed impact parameter $b$; indeed, it seems natural
to assume that events with small $b$ usually have larger  $n_{\rm jet \
pair}$. This leads to the eikonal formalism mentioned above.

Convenient and elegant, the eikonal method  reduces the rise of this 
cross-section, and allows one to enforce the requirement of s-channel 
unitarity.
Here one obtains the total cross-section through the eikonal formula
\begin{equation}
\sigma_{tot}=2\int d^2{\vec b}[1-e^{-\Im m\chi(b,s)}cos{\Re e\chi(b,s)}]
\label{stot}
\end{equation}
and one introduces the jet cross-section as the term which drives the 
rise in the eikonal function. This can be done unambiguously  by 
defining the inelastic cross-section\cite{rohini} given in  the eikonal 
formulation by
\begin{equation}
\sigma_{inel}=\int d^2{\vec b}[1-e^{-2 \Im m{\chi}(b,s)}] .
\end{equation}
This expression can also be obtained upon summing multiple 
collisions which are Poisson distributed with an average number 
$n(b,s)=2\ \Im m~{\chi}(b,s)$. 
Making the 
%further 
approximation $\Re e \chi=0$, one obtains a very simple
expression
\begin{equation}
\sigma_{tot}=2 \int d^2{\vec b}[1-e^{-n(b,s)/2}]
\end{equation}
%{\it
To proceed further, one needs to introduce the soft processes, which cannot be
described by perturbative QCD.
Following the separation shown in Sect.II, one can 
 approximate 
%}
 $n(b,s)$ 
by introducing a
%n {\it ad hoc} 
separation between soft and hard processes as:
%- as explained in the introduction - by different $b$-distributions
%between soft and hard processes 
\begin{equation}
n(b,s)=n_{soft}+n_{hard}=A_{soft}(b)\sigma_{soft}(s)+A_{jet}(b)\sigma_{jet}(s)
\label{nsplit}
\end{equation}
The separation between hard and soft processes is of course approximate
and so is the factorization into energy and transverse dimension dependence. 
However, it is useful as it allows one to break down the calculation into 
building blocks, which can be separately understood and  put together again 
later as  part of the whole structure.

The  procedure of Eq.(\ref{stot}) reduces the fast rise due to the mini-jet 
cross-section, but the extent to which this softens the rise, is highly
dependent on the impact parameter ($b$)-dependence of $n(b,s)$.
The simplest \cite{durand} ansatz, 
%{\it 
which introduces a minimum number of parameters
%} 
is to assume that the 
$b$-dependence is the same for 
both the soft and the jet component, and further that it is given by the 
Fourier transform of the electromagnetic form factor of the colliding hadron,
%colliding hadron electromagnetic form factor 
${\cal F}(q)$\cite{durand}. Thus for protons and antiprotons one will 
have
\begin{equation}
A_{soft}(b)=A_{jet}(b)={{1}\over{(2\pi)^2}}\int d^2{\vec b} 
e^{i q\cdot b} [{\cal F}_p(q) ]^2
={{1}\over{(2\pi)^2}}\int d^2{\vec b} 
e^{i q\cdot b} [{{\nu^2}\over{q^2+\nu^2}}]^4
\label{FF}
\end{equation}
%In these Eikonal Minijet Models (EMM), one still faces the 
%problem that the mini-jet driven cross-section
%is unable to reproduce  properly the rise, from the beginning up to 
%asymptotia,
%without introducing further parameters.
%{\it 
It is well known \cite{block} that 
%} 
these 
Eikonal Minijet Models (EMM) are unable to properly reproduce 
the experimentally observed, complete rise of the cross-section from the 
beginning up to the asymptotia, without introducing further parameters.
As an example, results obtained using Eq.(\ref{FF}) with $\nu=0.71\ 
GeV^2$, and current Gl\"uck, Reya and Vogt (GRV) parton densities for 
the proton~\cite{GRV} to 
calculate the jet cross-sections in Eq.(\ref{sigjet}), are shown  
\begin{figure}[htbp!]
%\begin{center}
%\epsfig{file=emm.eps,width=10cm}
\includegraphics[width=10cm] 
%{/afs/lnf/project/teorico/user/pancheri/sigtot/reggefig2005/emm05.eps}
{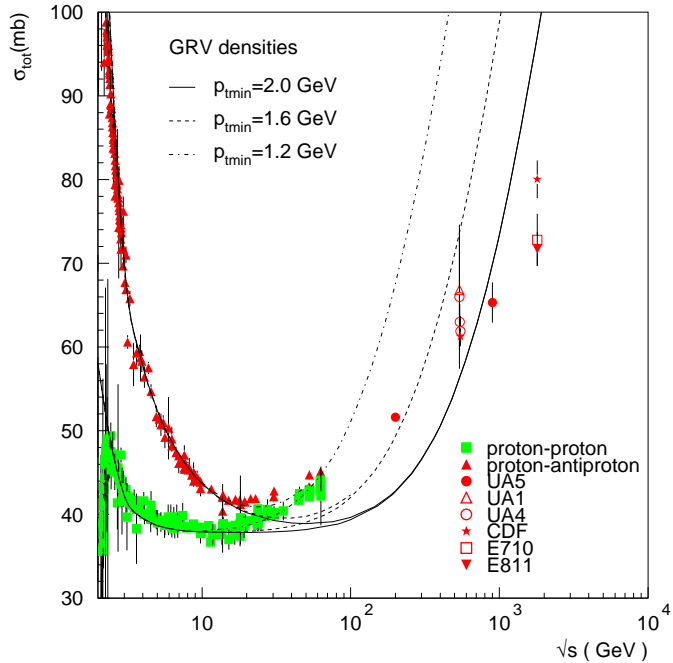}
\caption{\label{EMM}
Comparison between data \cite{PDG05,UA4,UA1,UA5,E710,CDF,E811} and  the EMM (see text),
for different minimum jet transverse momentum.}
%\end{center}
\end{figure}
in FIG.~(\ref{EMM}). One can see from the figure that a 
$p_{tmin}\approx  2\ GeV$ is needed, in order to obtain a numerical value of 
the total proton cross-sections in the 80 mb range, at Tevatron energies.
However, for such $p_{tmin}=2\ GeV$, the rise does not begin
until $\sqrt{s}$ is already in the 100 GeV region.
On the other hand, a smaller value for the regulator $p_{tmin}$, typically 
just above 1 GeV, would allow for the beginning of the rise around
$20\div 30\ GeV$, as the data indicate, but then the cross-section rises 
too rapidly in comparison with the Tevatron data.  In FIG.~(\ref{EMM}), in all 
jet cross-sections  computed using Eq.~(\ref{sigjet}),  we use  
the strong coupling constant $\alpha_s$ at scale  $p_{t}$. Even at 
$p_t=p_{tmin}$,  this value for the scale, albeit low, is still in
a range where the asymptotic freedom expression is approximately valid. For 
the low energy behaviour, all the curves shown in FIG.~\ref{EMM} are obtained 
with the same  phenomenological fit as in ref.\cite{ff2}.

%{\it 
Different solutions to the above problems have been adopted, 
including energy dependence of   the cut-off parameter
  $p_{tmin}$, adding more terms in Eq.~(\ref{nsplit}) or  using  Eq.~(\ref{FF})\cite{durand,block}  with different constants for the low and the 
high energy part.

 We have taken the approach  to keep fixed
$p_{tmin}$ (for a given beam and target combination) and have an energy
dependent overlap function, the energy dependence being modelled by a QCD
motivated calculation. The QCD motivated model gives a transverse momentum 
distribution of the partons which is energy dependent and this, in turn,
makes our overlap functions energy dependent.
We shall see in the next sections that 
%We are reluctant at this stage to dismiss the fixed $p_{tmin}$ approach
%because we wish to show the effect that 
in this model soft gluon emission has  an effect very similar to that of  
an energy dependent $p_{tmin}$.  
%}

%\end{document}

\section{
%Analiticity 
Analyticity requirements on the impact parameter distribution}
Within the context of QCD, the  problem described in the previous
section, sometimes referred to as the soft and hard Pomeron problem,
can have (at least) two different origins. If 
complete factorisation between the dependence on the energy $\sqrt{s}$ and
the impact parameter $\vec{b}$ holds, then the abovementioned  soft and 
hard pomeron
prolem may simply be taken to be indicative that the s-dependence generated 
through gluon densities, is wrong.
%complete factorization between
% energy dependence from the impact parameter distribution holds, 
%(as in Eq.(\ref{nsplit})), then it may simply be that the s-dependence, 
%as generated through the gluon densities, is wrong. 
On the other hand, given the fact that gluon densities are measured in 
Deep Inelastic Scattering  experiments, one may think of an alternative  
explanation
of the probelm. This second,
complementary explanation is that not all the s-dependence is due to the 
jet-cross-sections and  there is further energy dependence in 
the impact parameter distribution. Some general analyticity arguments 
can be invoked to shed light on the above and thus limit the 
arbitrariness in the choice of the function $\chi(b,s)$.

Consider the elastic scattering amplitude $T_{el}(s,t)$ 
(with $s$ and $t$ the usual Mandelstam variables) normalized so 
that the total cross section is given by
\begin{equation}
\label{sigtot}
\sigma_{tot}=2 \int d^2{\vec b}[1 - e^{-\Im m~\chi(b,s)} 
\cos\Re e~\chi(b,s)]= ({{2}\over{s}})\ \Im m~{T_{el}}(s,t=0),
\end{equation}
and the elastic differential cross section by
\begin{equation}
{{d^2 \sigma_{el}}\over{d^2{\vec q}}}={{1}\over{4 \pi^2 s^2}}|T_{el}|^2
\end{equation}
Consistently with Eq.(\ref{sigtot}), one then has
\begin{equation}
T_{el}(s,t)=
i s\int d^2{\vec b}e^{i{\vec q}\cdot {\vec b}}[1-e^{i\chi(b,s)}].
\end{equation}
For complete absorption, i.e. $\Re e~{\chi}=0$, we get 
\begin{equation}
T_{el}(s,t)=i s\int d^2{\vec b}e^{i{\vec q}\cdot {\vec b}}[1-e^{-n(b,s)/2}]
\end{equation}
%and the total cross-section reduces to
%

Let us briefly examine restrictions imposed on the large b-behaviour of the 
function $n(b,s)$ by the requirements of analyticity. Consider the Fourier 
transform of the elastic scattering amplitude, 
\begin{equation}
\label{Ampl}
{\cal A}(s, b)=
{{-i}\over{4\pi s}}\int_{-\infty}^0 dt\ T_{el}(s,t) J_0(b\sqrt{-t})
\end{equation}
where $t=-{\vec q}^2$, with ${\vec q}$ the transverse momentum variable.
Equivalently, we have
\begin{equation}
T_{el}(s,t)=\pi i s \int_0^{\infty} d b^2 J_0(b\sqrt{-t}) {\cal A}(s,b)
\end{equation}
The finite range of hadronic interactions implies that the partial wave 
expansion converges beyond the physical region, i.e., throughout the 
Lehmann ellipse. This requires that $T_{el}(s,t)$ be analytic in $t$ 
up to $t=4\mu^2$, where $\mu$ is the pion mass. For positive $t$, 
we continue the above expression for $\sqrt{-t}=iW$, with $W$ real and 
positive. In this ``unphysical'' region, we have
\begin{equation}
T_{el}(s,W^2)=\pi i s \int db^2 I_0(bW) {\cal A}(s,b^2)
\end{equation}
For large b, $I_0(bW)\sim {{e^{bW}}\over{\sqrt{2\pi}bW}}$, so that 
for the integral to converge, one needs
\begin{equation}
|{\cal A}(s,b^2)| < e^{-bW_o}\ \ \ \ with  \ \ \ W_o\simeq 2 \mu
\end{equation} 
In the purely absorptive model $(\Re e~\chi = 0)$ thus,
\begin{equation}
1-e^{-n(b,s)/2} <e^{-bW_o/2},
\end{equation}
and we see that $n(b,s)$ must be bounded at least by an exponential. 
Stronger (but model dependent) constraints arise provided one imposes 
that the elastic differential cross-section exhibit a ``diffraction peak''. 
That is
\begin{equation}
T_{el}(s,t)\simeq f(s) e^{{\hat b}(s) t}
\end{equation}
where ${\hat b}(s)$ is the so-called width of the diffraction peak, which
has an observed ( approximately) logarithmic s dependence. Then, 
Eq.(\ref{Ampl}) gives
\begin{equation}
{\hat {\cal A}}(s,b)
\simeq {{i f(s)}\over{4\pi s}}
 \int dq^2 J_0(b q)
e^{-{\hat b}(s)q^2}
={{i f(s)}\over{2\pi s}}
\left[ {{1}\over{2 {\hat b}(s)}} \right] 
e^{-{{b^2}\over{4{\hat b}(s)}}},
\end{equation}
which requires a Gaussian fall-off of the amplitude
in the impact parameter $b$, with its scale determined by the width of the 
diffraction peak. In the Regge pole description, 
\begin{equation}
{\hat b}(s) \sim \alpha' ln(s/s_0), \ \ \ \ \ \ \ \ \ f(s)\sim -i\beta
(s/s_o)^{1 + \epsilon}
\end{equation} 
and 
\begin{equation}
{\cal A}(s,b)
\simeq 
{{\beta(s/s_o)^{\epsilon}}\over{4 \pi (\alpha's_o) ln(s/s_0)}} 
e^{-
{{b^2}\over{4  \alpha' ln(s/s_0)}} }
\end{equation}

In the EMM, where the impact parameter distribution is given by the 
Fourier transform of the proton form factor, a model we refer to as
the Form Factor (FF) model, one has
\begin{equation}
n(b,s)={{\nu^2}\over{96 \pi}}
 \left(\nu b \right)^3 K_3(\nu b)[\sigma_{soft}+\sigma_{jet}]
\end{equation}
The modified Bessel functions of the third kind $K_{\mu}(z)$ are
bounded by an exponential at large values of the argument, i.e.
$K_{\mu}(z) \sim \sqrt{{\pi}\over{2z}} e^{-z} \{1 +{\cal O}
({{1}\over{z}})\}$. We see then that the EMM in the FF formulation does 
satisfy the requirements of analyticity in the Lehmann ellipse.
But in the EMM, the observed shrinking of the diffraction peak, 
corresponding to an energy dependent near gaussian fall off, 
is not present. Even if one were to introduce an {\it ad hoc } 
energy dependence (as is often the practice) instead of the
constant scale parameter $\nu$, as in the FF model, still one 
would be nowhere near the stronger Gaussian decrease at large 
impact parameter values. This is one reason why the FF model 
in the eikonal formulation, where jet cross-sections  drive 
the rise, fails to provide an adequate description of the overall 
energy dependence of the total cross section, without introducing an 
ad hoc modification of the scale parameters.
%{\it 
\section{ The soft gluon transverse momentum distribution}
The model for the total cross-section presented in this paper is based on
the ansatz that QCD provides the main processes  at work leading to the
observed energy rise of all measured total cross-sections. An important
motivation of this model is to make quantitative 
calculations
based on current QCD phenomenology, namely current parton densities with their
energy momentum dependence, running behaviour of the coupling constant and
soft gluon resummation techniques. Of all these at present soft gluon 
resummation is the one which presents the toughest technical challenge.
% And, in this respect,
%the harder to implement as well. 

Let us begin by considering the well known function which describes soft 
gluon emission from a parton-parton pair, namely the soft gluon transverse 
momentum distribution\cite{ourPR,DDT} 

\begin{equation}
{{
d^2P({\bf K_\perp})
}
\over{
d^2{\bf K_\perp}
}}
\equiv 
\Pi(K_\perp)=\int {{d^2\vec{b}}\over{(2\pi)^2}} e^
{i
{\bf K_\perp \cdot b}
-h(b)
}
\end{equation}
with
\begin{equation}
h(b)=\int d^3{\bar n}_g(k)[1-e^{-i{\bf k_\perp \cdot b}}]=\int
{{d^3 k}\over {2 k_0}} \sum_{i,j=colors}|j^{\mu,i}(k)j_{\mu,j}(k)|
[1-e^{-i{\bf k_\perp \cdot b}}]
\label{hbtrue}
\end{equation}
where $d^3{\bar n}_g(k)$ is the  distribution for single gluon emission in
a scattering process, and $j^{\mu,i}$ the QCD current responsible for 
emission. The above expression has been widely used to study the 
initial state transverse momentum distribution in Drell-Yan processes
\cite{PP,greco} as well as  W-production \cite{halzenscott,mario}. 
%It is also presently 
%used for the energy dependence of Generalized Parton 
%Distributions \cite{sterman}. 

In the limit of large $k_\perp b$,  one can neglect the
exponential term, and the above expression 
%can be evaluated 
%between  the upper limit $Q$ and a lower one set to be the QCD constant %$\Lambda$, and it
 reduces to the well
known Sudakov form factor \cite{Sudakov}, namely
\begin{equation}
h(b)\approx S(b)=
\int 
d^3{\bar n}_g(k)
=\int
{{
d^3k}\over{2 k_0
}} \sum_{colors}
|j^{\mu,i}(k)j_{\mu,j}(k)|
\end{equation}
Introducing the running coupling constant and its asymptotic freedom
expression, the integration from $1/b$ up to  an upper limit $Q$ gives \cite{sterman}
\begin{equation}
h(b) \approx {4 C_F\over(11-2n_f/3)} ln{Q^2\over \Lambda^2} 
ln{{ln (Q^2/\Lambda^2)}\over {ln (1/b^2\Lambda^2)}}
\end{equation}

For small momenta, 
however,  the above expression is not sufficient to   reproduce the 
observed  tranverse momentum distribution  in various hadronic processes.
% one introduces 
%and a cut-off is
%introduced, which 
Thus, an intrinsic transverse momentum (of Drell-Yan pairs or W-boson or partons, depending on the physical process under
consideration) has to be  introduced. The function now reads \cite{jk}
\begin{equation}
h(b)=b^2p_{\perp int}^2 + S(b)
\end{equation}
where the intrinsic transverse momentum $p_{\perp int}$ is a constant, of the order of a few 100 MeV, parametrized according to the process under consideration.

In our model,  the function in Eq.~(\ref{hbtrue}), in addition to describing various hadronic transverse momentum effects, also plays    a major role in total cross-section
calculations. Our model for the hadronic transverse momentum distributions due to soft gluon resummation differs in two major points from what we have just described. Basically,  we focus   
%to  study  Eq.~(\ref{hb})  literally, both at
on  the lower and higher limits of integration in Eq.~(\ref{hbtrue}). 
At the lower limit,  
since this expression refers
to soft gluons, we suggest that the correct use of this equation requires to integrate the gluon momentum 
down to $k_\perp=0$ and avoid the introduction of {\it ad hoc} quantities such as  the intrinsic transverse momentum.  At the upper limit, one needs to specify, for each
given process, how the maximum soft gluon energy is defined. However, 
in order to use the above expression for a believable calculation, a number of 
points need to be clarified, namely
\begin{description}
\item(i) whether it makes sense to   use a parton picture when the gluon
 momenta become close to zero, with the related question of  
 what is the behaviour  of the strong 
coupling constant $\alpha_s$ when
one integrates the gluon momentum  down to zero 
\item(ii) whether the emitting particles are quarks or gluons
\item(iii) what are  the constraints from kinematics upon the maximum gluon
momentum
\end{description} 
We shall discuss some of these issues in detail in the sections to come, here
we comment briefly on these three points.

Concerning the parton picture,
%It is important to notice that, 
while we use it 
%the parton description 
for the mini-jet contribution to the cross-section,   soft gluon Bloch-Nordsieck resummation  factors out of the LO basic scattering process, and is thus independent of parton densities,   involving only the momenta 
%of the scattering partons 
and the QCD coupling between soft gluons and  the emitting partons. On the 
other hand  when $k_\perp \rightarrow 0$, this coupling is not an observable 
quantity, since it refers to a single soft gluon emission, and  one soft 
gluon is not an observable quantity (only its integrated spectrum is). 
%Our main point, in this and in previous papers, is that in order to take %soft gluons in proper consideration,
% their contribution needs to be resummed, because one soft gluon is not an %observable,  and integrated over all momenta, starting from zero. 
%, because hence an exponentation of the relevant integrated spectrum. %This
%resummation can be derived also perturbatively, but we must notice that %it
%will imply an integration  down to zero of the  
%transverse momentum of the single gluons. 
This has two effects: firstly, one needs to use a non-perturbative 
expression for the QCD coupling constant, since the momenta
are so small, and 
%then  that what will matter in all these calculations is
%the {\it integral} of momenta of $\alpha_s.
secondly only the {\it integral} of moments of $\alpha_s$ will matter. 
The infrared behaviour of
$\alpha_s$ is a matter of speculation. We propose our own model, whose  justification rests upon a Regge description and on 
the Richardson type potential for quarkonium. As will be described in the next sections,   a specific form for  $\alpha_s$ is
chosen, singular in the infrared limit, but integrable \cite{ff2}.

Another issue to address relates to  the effect of emission from quarks, 
valence and sea, and from gluons in the parton processes. 
In this paper, we only deal with  emission from the initial
valence quarks, and use the relevant kinematics with their averages. This 
approximation is  justified by the fact that, relative to the emission from
the initial valence quarks,  emission from the gluon legs is
to be considered as emission from internal legs, thus subleading in
infrared terms. A complete calculation should of course include also
emission from partons other than the valence quarks and hence mostly 
the low-x gluons. We expect this inclusion may eventually increase the softening effect. 
%On the other hand the precision of this calculation is not yet sufficient %to realistically include such non leading effects. 
%, but we shall examine this contribution in a subsequent 
%paper, as here we are mostly interested in whether soft gluons can %produce
%the necessary softening effect on the rise of QCD mini-jets.

To describe  the softening effect quantitatively,
% and still work with present parton densities (where the transverse %momentum of the soft gluons is integrated over) 
one needs to  focus on the maximum transverse momentum allowed to single
gluon emission. 
%(the upper limit in the integral defining h(b,s))
This
quantity is energy dependent, as one can easily see using the kinematics of
single gluon emission in parton-parton scattering of initial c.m. energy $\sqrt{\hat s}$. For the   process 
\begin{equation}
p_i\ + \ p_j \rightarrow gluon \ +\ X
\end{equation}
where ${\hat s}=(p_i+p_j)^2$  and X a final state of given momentum Q, the maximum
transverse momentum of the gluon is given by
\begin{equation}
q_{max}(\hat s)={\sqrt{\hat s}\over 2}(1-{Q^2\over \hat s})
\label{qkin}
\end{equation}
If we consider the state X to be the final (mini) jet-jet system in the 
inelastic collision contributing to the cross-section,
%$n(b,s)$, 
then the above expression depends on the parton sub-energies and on the 
final state momentum of the jet-jet system, characterized by 
a transverse momentum $p_t \ge p_{tmin}$, where $p_{tmin}$ is a scale
chosen to separate hard and soft processes. In principle, for each subprocess of given ${\hat s}$ and $Q^2$ one should evaluate the function $h(b)$ with the above $q_{max}$. In practice,  we use a value for the maximum transverse
momentum allowed to single gluons, which is averaged over the initial and
final parton momenta. This is shown more explicitely in \cite{ff2} and in section VI, but
the result is that for a given $p_{tmin}$ cut-off in the minijet
distribution, for the valence quarks,
 the scale $q_{max}$ increases with the c.m. energy of the
hadron-hadron system. This can be qualitatively understood by considering that  the valence quarks will on the average carry a larger energy and can then shed 
more soft gluons. 
%This is so because, for any given $p_{tmin}$, 
%on the average there will more partons of higher energy participating to the
%collisions, hence more momentum can go out in soft gluons. 
 Thus, in the picture we present for  hadron-hadron collisions,  as the overal
energy increases, we have more parton-parton collisions for the same $p_{tmin}$
(since the number of low x-gluons increases) but also more energy available 
to soft gluons both from initial valence quarks and from all the hard partons 
in general, and thus more of a reduction. It is the balancing of these two 
effects which we believe to be responsible for the observed softer rise of 
total cross-sections.
%}

%\end{document}
\section{Soft Gluon emission and energy dependence in the impact parameter
distribution}
In our previous work, we have advocated that a cure for the difficulty 
in obtaining the early dramatic rise and the softer asymptotic behaviour
{\it simultaneously}, lies within QCD itself; viz., the ubiquitous soft 
gluon emission accompanying all QCD scatterings which can slow down any 
abrupt rise in the cross-section.  To make this quantitative, we put forward 
a model for the impact parameter distribution of partons in the hadrons,
based on the Fourier transform of the transverse momentum distribution of the
soft gluons emitted in the collisions, as described in the 
Block-Nordsieck (BN) summation procedure.This distribution is energy 
dependent  simply because 
%it depends upon 
the maximum energy allowed to each single emitted soft gluon, 
%which  
in turn depends on the energy of the colliding partons.
In detail, we have a picture of parton-parton
collisions at all admissible subenergy values and
with a given transverse momentum due to initial
state radiation.  In 
our model the soft gluon resummed transverse momentum distribution of  
partons in the hadrons and the parton distribution in  impact parameter 
space are Fourier Transforms of each other. 
In principle, this formalism could be used 
to obtain impact parameter dependent parton densities, but our aim 
in the present paper  is to obtain a
prediction for the  total cross-section based on currently used QCD
functions and parameters such as parton densities and $\Lambda_{QCD}$. 
We thus follow our previous proposal \cite{ff2,ff} to average out the
behaviour of partons in their transverse momentum variable and 
arrive at the following expression
\begin{equation}
2\Im m~\chi(b,s)=n(b,s;q_{max},p_{tmin})=n_{soft}+
n_{jet}=A_{soft}(b)\sigma_{soft}+
A_{BN}(b,q_{max})\sigma_{jet}
\end{equation}
As mentioned earlier, the eikonal formulation provides a 
natural framework, in which different contributions to the
total cross-section can be resolved into their 
various structural elements: the rise is 
incorporated in $n_{jet}$, and the decrease and normalization 
in $n_{soft}$. In our previous work, we had parametrized 
phenomenologically the soft part, and used perturbative QCD  
for the jet part. In this paper, we study whether
soft gluon summation can describe  the (experimentally observed) initial 
decrease in proton-proton scattering.

For this purpose, we write
\begin{equation}
n(b,s)=A_{BN}^{soft}\sigma_{soft}+A_{BN}^{jet}\sigma_{jet}
\end{equation}
with
\begin{equation}
\label{abn}
A_{BN}={{e^{-h(b,s)}}\over{\int d^2{\vec b}
e^{-h(b,s)}}}
\end{equation}
where from Eq.(\ref{hbtrue}) we have
\begin{equation}
\label{hb}
h(b,s)={{8}\over{3\pi}}\int_0^{q_{max}}{{dk}
\over{k}}
\alpha_s(k^2)\ln({{q_{max}+\sqrt{q_{max}^2-k^2}}
\over{q_{max}-\sqrt{q_{max}^2-k^2}}})[1-J_0(kb)]
\end{equation}
and $q_{max}$ 
%is a function of energy, which
depends on energy and the kinematics of the process\cite{greco}. 
From Eq.(\ref{qkin}), the
following average expression for $q_{max}$ was proposed in our
previous paper \cite{ff2},
\begin{equation}
\label{qmaxav}
M\equiv <q_{max}(s)>={{\sqrt{s}} 
\over{2}}{{ \sum_{i,j}\int {{dx_1}\over{ x_1}}
f_{i/a}(x_1)\int {{dx_2}\over{x_2}}f_{j/b}(x_2)\sqrt{x_1x_2} \int_{z_{min}}^1
 dz (1 - z)}
\over{\sum_{i,j}\int {dx_1\over x_1}
f_{i/a}(x_1)\int {{dx_2}\over{x_2}}f_{j/b}(x_2) \int_{z_{min}}^1 (dz)}}
\end{equation}
with $z_{min}=4p_{tmin}^2/(sx_1x_2)$ and $f_{i/a}$ the valence 
quark densities used in the jet cross-section calculation.

M establishes the scale which, on the average, regulates soft 
gluon emission in the collisions, whereas $p_{tmin}$ provides the scale 
which characterizes the onset of hard parton-parton scattering. 
For any parton parton subprocess characterized by a $p_{tmin} \approx 
1\div 2\ GeV$, M has a logarithmic  increase at 
reasonably low energy and  an almost constant behaviour 
at high energy\cite{ff2}. 
The eikonal formalism which we use to describe the total
cross-section, incorporates
multiple parton parton collisions, accompanied by
soft gluon emission from the initial valence
quarks, to leading order. Notice that in this
model, we consider emissions only from the
external quark legs. In  the impulse approximation on 
which the parton model itself is based,  the valence quarks are
free, external particles. In this picture, emission of soft 
gluons from the gluons involved in the  hard  scattering, is non leading.
As the energy increases, more and more hard
gluons are emitted but  there is also a 
larger and larger probability of soft gluon emission : the 
overall effect is a rise of the cross-section, tempered  
by the soft emission, i.e. the violent mini-jet rise due to 
semi-hard gluon gluon collisions is tamed by soft gluons. Crucial 
in this model, are the scale and the behaviour of the strong 
coupling constant which is present in the integral over the soft 
gluon spectrum. While in the jet cross-section $\alpha_s$ never plunges into
the infrared region, as the scattering partons are by construction semi-hard, 
in the soft gluon spectrum the opposite is true and a regularization is 
mandatory. We notice however that here, as in other  problems of soft hadron 
physics\cite{doksh}, what matters most is not the value of $\alpha_s(0)$, 
but rather its integral.  Thus, all that we need to demand, is that $\alpha_s$ 
be integrable, even if singular~\cite{nak}. We employ the same 
phenomenological expression for $\alpha_s$ as used in our previous works, 
namely
\begin{equation}
\label{alphaRich}
\alpha_s(k_\perp)={{12 \pi }\over{(33-2N_f)}}{{p}\over{\ln[1+p({{k_\perp}
\over{\Lambda}})^{2p}]}}
\end{equation}
Through the above, we were able to reproduce the effect of the
phenomenologically introduced intrinsic transverse momentum of 
hadrons~\cite{nak}, and more recently obtained a very good 
description of the entire region where the total cross-section 
rises\cite{ff2}. 
This expression for $\alpha_s$ coincides with the usual one-loop expression 
for large values of $k_\perp$, while going to a singular limit for small 
$k_\perp$. For $p=1$ this expression corresponds to the Richardson
potential\cite{richardson} used in bound state problems. We see 
from Eq.(\ref{hb}) that $p\ =\ 1$, leads to a divergent integral, and thus 
cannot be used. Notice that, presently, in the  expression for $h(b,s)$, 
the masses of the emitting particles are put to zero as is
usual in perturbative QCD. Thus, for a convergent integral,
 one requires $p<1$ and the successful 
phenomenology indicated in \cite{ff2} gave $p=3/4$. However, 
more study is needed, especially in the full utilization of the 
Bloch-Nordsieck description, before one can completely define 
%the actual 
an expression for $\alpha_s$ in the infrared limit. 
A different possibility is to use a so-called frozen $\alpha_s$ model, 
 for which 
$\alpha_s(k_\perp^2)={{12\pi}\over{27 
\ln[(k_\perp^2+a^2\Lambda^2)/\Lambda^2)]}}$.
These two expressions lead to very different large-b behaviour of the
function $n(b,s)$ and,
in light of the above discussion concerning the shrinking of the
diffraction peak, give quite a different s-dependence in the rising region 
of the  total proton cross-section.

In \cite{ff2}, we presented analytic approximations
for the function 
$h(b,s)$, obtaining, in the frozen $\alpha_s$ case
\begin{equation}
\label{halphafroz3}
\lim_{b\rightarrow \infty} h(b,M,\Lambda) ={{2c_F {\bar \alpha_s}}
\over{\pi}}
\Biggl [
{{1}\over{4}}\ln(2Mb)+ 2 \ln(Mb)\ln(a{\Lambda}b)
-\ln^2{(a \Lambda b)}\Biggl ]\nonumber 
\end{equation} 
while for the singular case
\begin{eqnarray}
\label{halphas3}
\lim_{b\rightarrow \infty} h(b,M,\Lambda) 
&=& {{2c_F{\bar b}}\over{\pi}} (b^2\Lambda^2)^p\Biggl [ 
{{1}\over{8(1-p)}} 
\left( 2\ln(2Mb)+{{1}\over{1-p}}\right) +\nonumber \\
 &\ &{{1}\over{2p}} \left(2\ln(Mb)-{{1}\over{p}}\right)
\Biggl ]\nonumber
\end{eqnarray}
where $M\equiv q_{max}$, ${\bar \alpha}_s={{12\pi}\over{27 \ln{a^2}}}$,
${\bar b}={{12\pi}\over{33-2N_f}}$. From the above, 
one can see that an approximately Gaussian limit results in
 the singular $\alpha_s$ case, but not in  the frozen case.
Indeed, for the singular case, one has, to the lowest order,
$\lim_{b\rightarrow \infty} n(b,s) \sim  e^{-b^{2p}}$ i.e. an
exact Gaussian limit for the Richardson potential, which corresponds 
to $p=1$. This provides the theoretical reason why the entire region
where the total cross-section rises is well described by perturbative QCD 
(jet cross-section)  combined with
Bloch-Nordsieck summation with a singular $\alpha_s$ in the
infrared region. In contrast, neither the frozen $\alpha_s$ model 
nor, the Form Factor (FF)  model are successful there.

Other models for the behaviour of $\alpha_s$
in the infrared region, and studies of   the range of variability of
the parameters used in Eq(\ref{alphaRich})  will be
presented in a forthcoming publication.

\section{The decrease prior to the onset of mini-jets}
Let us now address the question of the lower energy range, 
prior to the rise, using the same phenomenological and theoretical 
tools of \cite{ff2}, but abandoning the FF and the frozen $\alpha_s$ 
models which appear to be inadequate.
To study the low energy region, we apply our
procedure to proton-proton scattering, where the
absence of resonances in the s-channel and leading
Regge poles in the t-channel make the picture
remarkably simple. At low energies, say before 10
GeV in the proton-proton c.m. system, one observes a very soft decrease,
which converts in a rise at an energy of $\approx $ 15 GeV in the c.m. 
In this low energy region, we know that gluon densities 
are still very small, and that (almost) all hard parton-parton scattering
takes place   among valence quarks : $\sigma_{jet}$, as defined through
Eq.(\ref{sigjet}), is a few thousands of the observed $\sigma_{tot}$.
In the break up of $\Im m\ \chi$ into a soft and a hard
part, the parameter $p_{tmin}$ separates hard and soft processes, namely
for $p_t^{parton}\ge p_{tmin}$ one counts parton-parton processes as
part of the jet cross-section, whereas for
$p_t^{parton}\le p_{tmin}$ the process can be counted as part of 
$\sigma_{soft}$.
Thus, in this region, we can study the contribution of
valence quark scattering without 
%the clouding of the
complications from inelastic gluon-gluon collisions.

  This region then exhibits the effect of soft 
gluon emission accompanying gluon exchanges among
the valence quarks. At higher energies, these
soft interactions 
%will 
still take place and be a substantial part of the cross-section, but they
will be shielded by the more dramatic behaviour of the perturbative QCD 
processes, since as the energy increases, smaller and smaller x-values of 
the gluon densities are probed and gluon exchanges among gluons start 
becoming important. Thus we must build a piece of the total cross-section 
which will survive at high energies, but which does not contribute
to the rise. To begin with, we start with a very simple ans\"atz :
that for proton-proton the cross-section 
$\sigma_{soft}$ is a constant and
the slight decrease comes from the straggling,
acollinearity effect of soft gluon emission.
We therefore
 propose, in first instance, the following expression for the
average number of soft collisions 
\begin{equation}
n_{soft}(b,s)=A_{BN}^{soft}\sigma_0
\end{equation}
with $A_{BN}^{soft}$ calculated through Eqs.(\ref{abn},\ref{hb}) 
 and investigate whether it is possible to find a constant
$\sigma_0$ and a set of parameters ($q_{max}$)  which
can describe $p p$ scattering at low energy. For the soft part,  the scale
$q_{max}$ corresponds to
the maximum energy allowed to soft gluons accompanying scattering with a
final parton transverse momentum smaller than $p_{tmin}$. We are dealing
with soft emission (for hard gluons  $p_t>p_{tmin}$) and thus we expect
$q_{max}$ not to be larger than $10-20$ \%  of $p_{tmin}$. This provides
an upper bound for $q_{max}$ for soft processes.

The observation is then that for processes contributing to $n_{soft}$, 
a soft gluon will always carry away less energy than for those 
contributing to 
$n_{hard}$. The question is how much lower is the allowed energy. We have 
proceeded phenomenologically, and found a set of values which, as will be 
shown in the last section, can give an acceptable description for 
$\sigma_{pp}$ before the rise.  These values are shown in Table 
\ref{qmaxtable}. Notice that, 
\begin{table}
\caption{\label{qmaxtable} Average $q_{max}$
 values used for the impact parameter distribution of the soft part of 
the eikonal}
\begin{tabular}{|c|c|}
\hline
$\sqrt{s}(GeV)$&$q_{max}^{soft}(GeV)$\\
5.&0.19\\
6.&0.21\\
7.&0.22\\
8.&0.23\\
9.&0.235\\
10.&0.24\\
50.&0.24\\
100&0.24\\ \hline
\end{tabular}
%\end{ruledtabular}
\end{table}
in order to reduce the number of free parameters, we assume that, 
at low energies, there is only one value of $q_{max}$, for both hard and soft
processes. However, as  the energy increases, the scale characterizing the 
soft processes does not grow as the one for the hard case,  
the latter obtained through the kinematics of jet production
in a hard parton parton scattering. Here we should
expect the scale to start as a slowly increasing
function of $s$, but to become a constant as
soon as hard processes become substantial for 
$\sqrt{s}\ge 10\ GeV$.
This is necessary, i.e. $q_{max}$ does not increase
indefinitely, because as the energy available to
soft gluons increases, at a certain point the soft
gluons will become hard and then undergo scattering among 
themselves.

Clearly, a soft scale not larger than 240 MeV is consistent
with our understanding of how a proton is structured, if
we attribute $\sigma_{soft}$, the soft component, as the cross section
when the scattering protons (and antiprotons) manifest themselves as
a quark and a spin zero diquark. The point is that the slopes for
meson and baryon Regge trajectories are justifiably equal only if 
a baryon is pictured as a quark/diquark system similar to a meson as a 
quark/antiquark system (thus making the string tensions equal). 
This mode for the nucleon is soft and diffusely spread over about a
fermi\cite{predazzi}. Consequently, the soft- gluon radiation distribution 
must be limited (lest it break the system). Thus, we estimate for the
soft process a $q_{max}\ \approx 1/(1~{\rm fermi} )\ =\ 0.2\ GeV$.

\begin{figure}[hbtp!]
%\begin{center}
\includegraphics[width=10cm]
%{/afs/lnf/project/teorico/user/pancheri/sigtot/regge/new_figs/qmax2004.eps}
{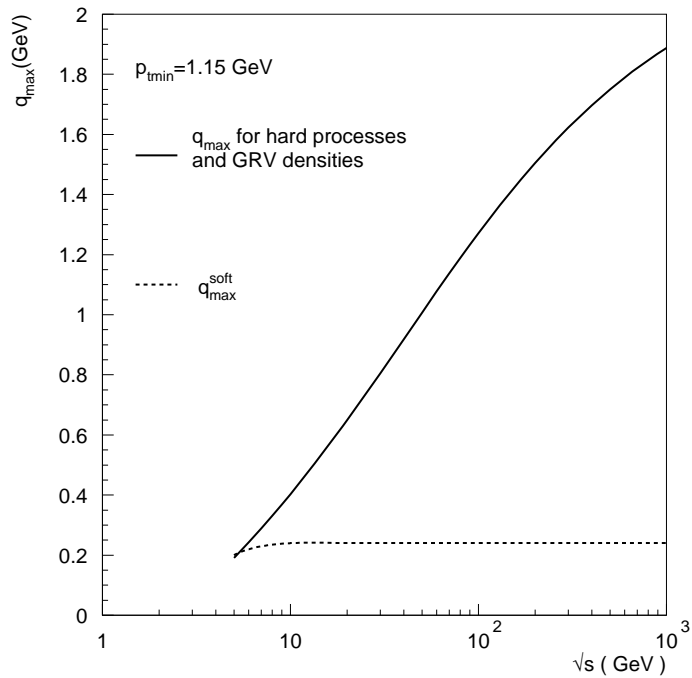}
\caption{\label{fig:qmax}The maximum transverse momentum allowed
(on the average) for single soft gluon emission
as a function of the c.m. energy of scattering
hadrons.}
%\end{center}
\end{figure}

With the value of $p_{tmin}$ which gave a smooth description of the total 
cross-section in \cite{ff2}, we plot in FIG.~(\ref{fig:qmax})
  the behaviour of 
$q_{max}$ as a function of energy, where the upper curve is the one obtained
using Eq.(\ref{qmaxav}), for $p_{tmin}=1.15\  GeV$, whereas the soft
 $q_{max}$ 
starts with the same value obtained for the jet term, and then is made to 
become a constant when it reaches 240 MeV.

With these values for $q_{max}$ we can now
calculate $A_{BN}$ for both the hard and soft 
terms in the eikonal
\begin{figure}[htbp!]
%\begin{center}
\includegraphics[width=10cm]
%{/afs/lnf/project/teorico/user/pancheri/sigtot/regge/new_figs/abn2004.eps}
{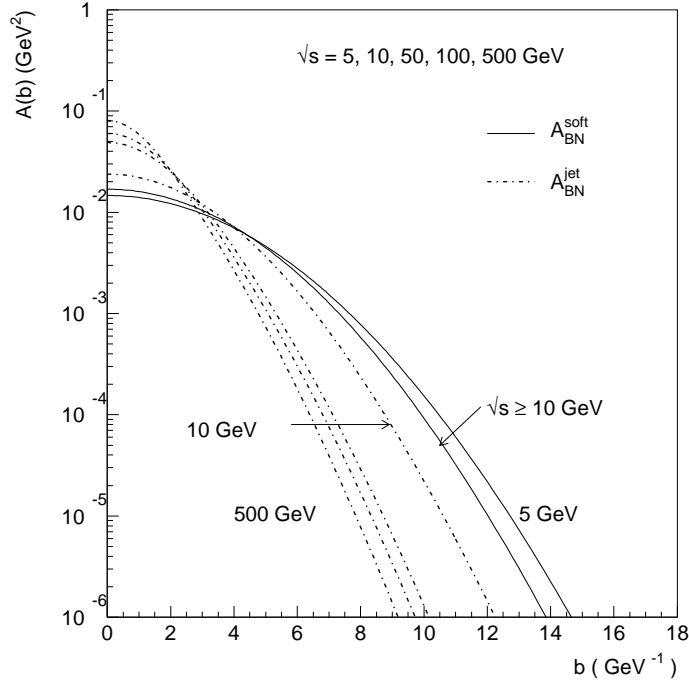}
\caption{\label{fig:ABN}The impact parameter distribution function 
calculated for the soft gluon summation model, using 
the $q_{max}$ values described in the text, for various energy 
values. Full lines are for the soft term.}
%\end{center}
\end{figure}
 as a function of the
impact parameter $b$.  Both soft and hard $A_{BN}$ are shown in
Fig.~\ref{fig:ABN} for a set of representative c.m. energies, 
$\sqrt{s}=5,10,50,100,500\ GeV$. Notice that  $A_{BN}^{soft}$ 
does not change for $\sqrt{s} \ge 10\ GeV$, 
since $q_{max}$ remains constant.

\section{Normalization and total cross-sections}
In order to obtain the average number of collisions and
thus the total cross-sections, the overall normalization,
given by $\sigma_{soft}$ has to be determined.

\begin{figure}[htbp!]
%\begin{center}
\includegraphics[width=10cm]
%{/afs/lnf/project/teorico/user/pancheri/sigtot/regge/new_figs/nbs2004.eps}
{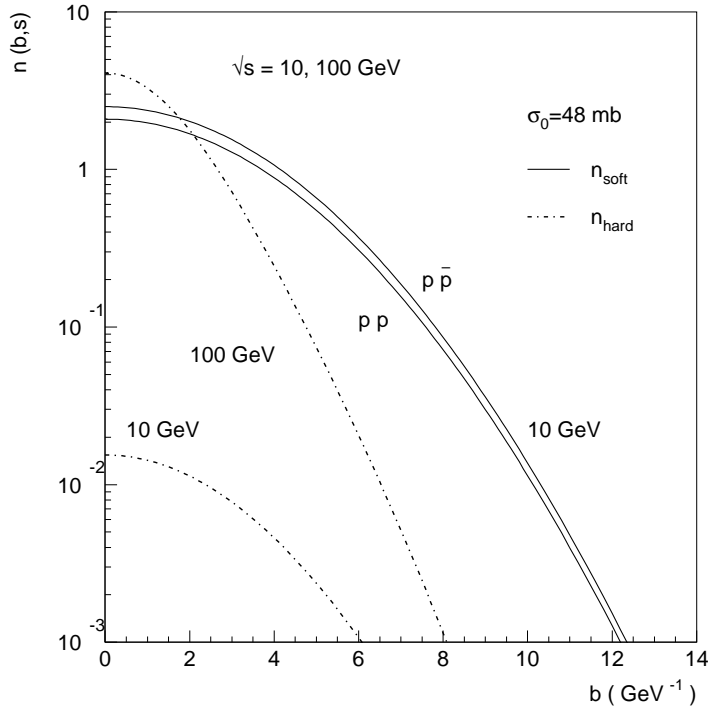}
\caption{\label{ncoll}
The average number of collisions  at $\sqrt{s}=10,100 GeV$ is plotted  
for the soft gluon summation model, using 
the $q_{max}$ values described in the text. 
The dot-dashed line corresponds to the jet 
contribution at $\sqrt{s}=10,100 \ GeV$. }
%\end{center}
\end{figure}

As mentioned in the introduction, we are 
%are considering a model in which 
assuming that the entire rise is due to $\sigma_{jet}$. For the 
proton-proton cross-section, one needs only one further parameter for 
the non perturbative region, namely a constant $\sigma_0$ which gives 
the normalization of the cross-section. As far as the proton-antiproton 
cross-section is concerned, 
%if 
the rapid decrease after the resonances is interpreted as dual to the 
Regge trajectory exchange and  it should be described by a power
$s^{\alpha_R(0)-1}\approx\ 1/ \sqrt{s}$ . Neglecting the real part 
of the eikonal, our model is now complete and reads as follows :
\begin{equation} 
\sigma_{tot}=2\int d^2{\vec b}[1-e^{-\Im m~\chi (b,s)}]
\end{equation}
with
\begin{equation}
2\Im m~\chi (b,s)=
A_{BN}(b,q_{max}^{soft})
 \sigma_{soft}^{pp,{\bar p}}+
A_{BN}(b,q_{max}^{jet})
 \sigma_{jet}(s;p_{tmin})
\end{equation}
We also have
\begin{equation}
\sigma_{soft}^{pp}=\sigma_0, \ \ \ \ \ \ \ \ \ \ \
\ 
\sigma_{soft}^{p{\bar p}}=\sigma_0
(1+{{2}\over{\sqrt{s}}}) 
\end{equation}
We find that, in order to properly reproduce the normalization of the
cross-section, we need a value  $\sigma_0=48\ mb$, in good agreement
with the considerations of Sect.II. We now show in FIG.~(\ref{ncoll}) 
the average number of collisions as a function of b, distinguishing 
between hard and soft contributions and using the values of 
$q_{max}$ shown in FIG.~(\ref{fig:qmax}).
 We only show the low energy region, 
$\sqrt{s}=10\div 100 \ GeV$ where the transition between soft and hard 
processes occurs.

\begin{figure}[htbp!]
%\begin{center}
\includegraphics[width=10cm]
%{/afs/lnf/project/teorico/user/pancheri/sigtot/reggefig2005/reggefig_2005.eps}
{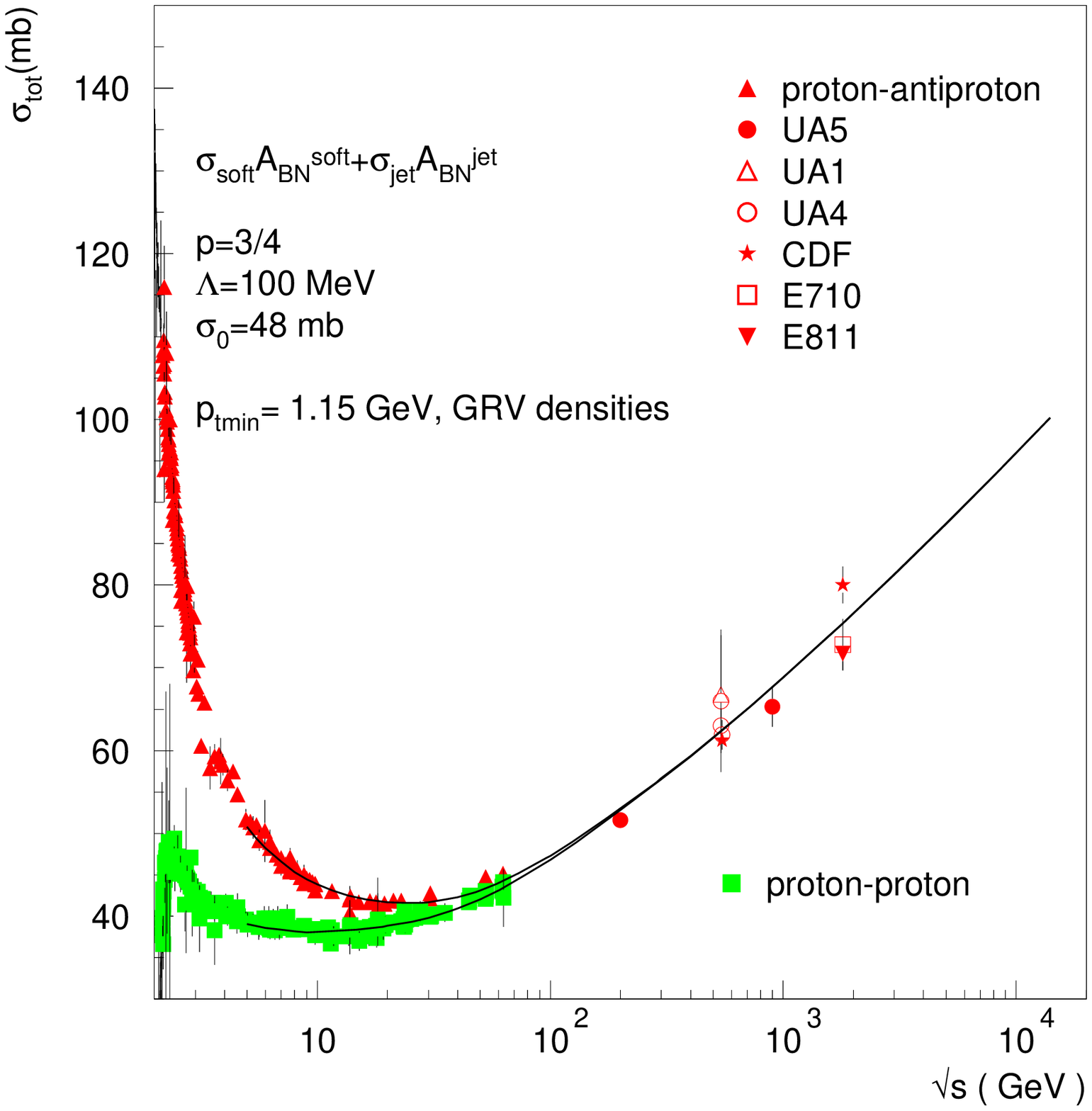}
\caption{\label{sigmas}Comparison of $p p$ and $p\bar{p}$ 
total cross-section data \cite{PDG05,UA4,UA1,UA5,E710,CDF,E811} 
 with results from the
%predictions in the 
Bloch-Nordsieck model described in the text.}
%\end{center}
\end{figure}
Finally, in FIG.~(\ref{sigmas}) we show the results
%predictions 
of our model, putting all the pieces together, for the total cross-section 
for proton-proton and proton-antiproton collisions. We see that the model 
gives an overall satisfactory description of the energy behaviour of
available data\cite{PDG05}.

%\end{document}

\section{Energy dependent $<b^2>$ in  the Bloch-Nordsieck model}
\begin{figure}[htbp!]
%\begin{center}
\includegraphics[width=10cm]
%{/afs/lnf/project/teorico/user/pancheri/sigtot/regge/new_figs/radius2004_rev.eps}
{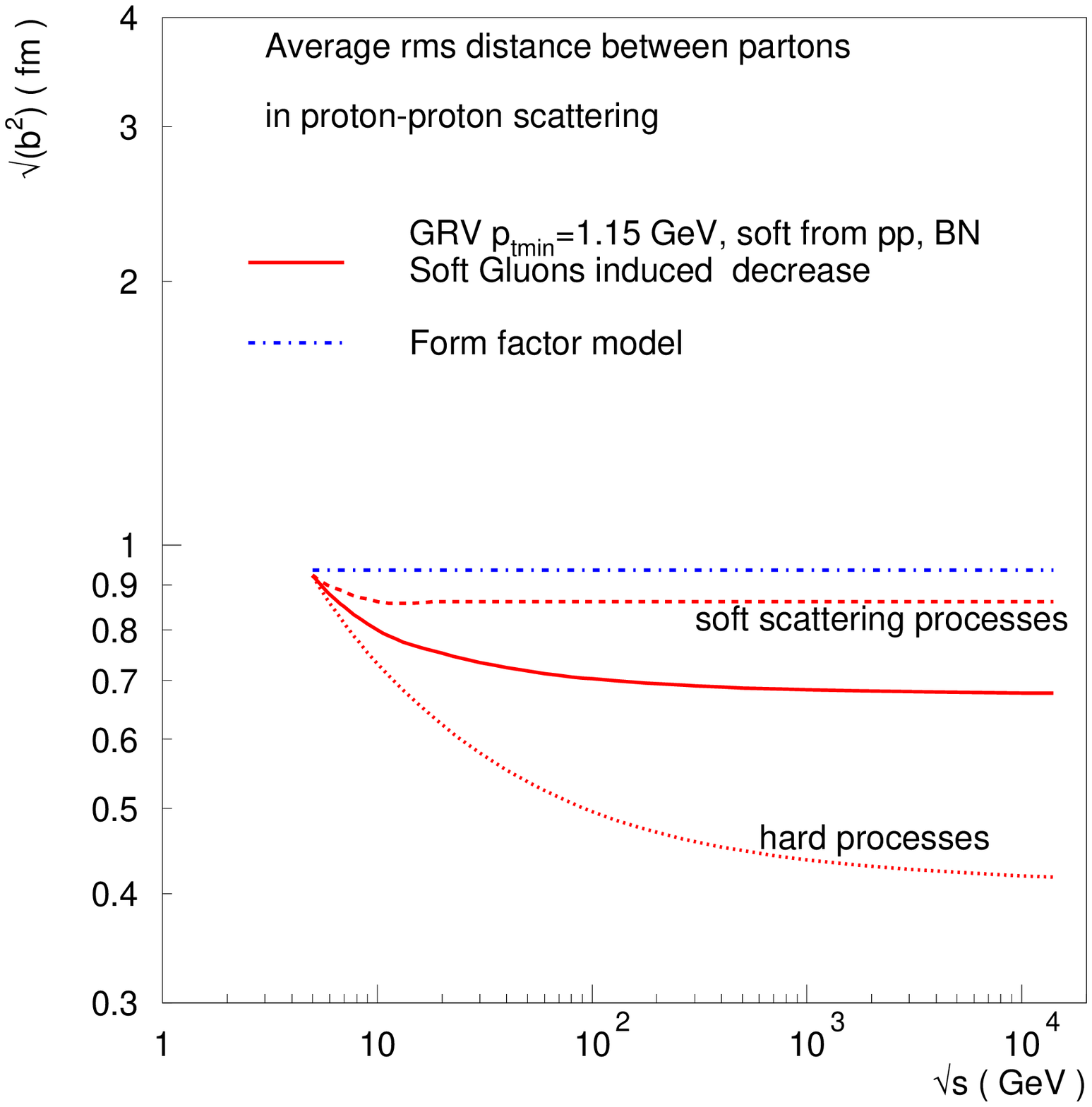}
\caption{\label{radius}The average distance between partons in the Bloch 
Nordsieck model (full
line) and the Form Factor model (dot-dashes) is shown as a function of
energy. For the former case, the contribution to this distance from the 
purely 
soft processes (dashes) with $p_t \le p_{tmin}$, and purely hard ones (dots) 
with  $p_t\ge p_{tmin}$, is also shown.}
\end{figure}

The energy dependent  transverse overlap function, $A_{BN}$,  discussed in the 
previous sections,  can be used to estimate the energy dependence of the 
average distance among partons in the transverse space during a scattering 
process, namely
\be
\label{b2}
<b^2>={{\int d^2 {\vec b}\  b^2 [A_{BN}(b,q^{soft}_{max}) +
A_{BN}(b,q^{jet}_{max})]}\over{\int d^2 {\vec b}  [A_{BN}(b,q^{soft}_{max})+
A_{BN}(b,q^{jet}_{max})]}}
\ee 
The energy dependence of the average rms distance between partons so defined, 
is shown in FIG.~(\ref{radius}).

One can see from this figure that for the hard part of the eikonal 
in the Bloch-Nordsieck  model (dotted curve), the mean distance between 
the scattering partons does decrease as the energy increases, thereby 
increasing the shadowing and taming the rise, as opposed to the Form 
Factor  model 
where $<b^2>$ is a constant. This is then further reflected in a more
modest high energy rise for the BN model as seen in FIG.~(\ref{sigmas}).
It is also pleasing to note the following self consistency. That is, 
at low  values of $\rs$, values of $<b^2>$ are the same for 
both the soft and hard part of the eikonal in both models, 
as they must since at low energy the transverse overlap function 
from the BN model is very similar to that from the FF model.

Observations about the need of a shrinkage in the radius of the proton,
have been made in Ref.~\cite{Borozan:2002fk,borozan-th}, where   
multiparticle production in hadron hadron interactions has been studied in 
detail in an eikonal Monte Carlo model. They find that in the hard
multi parton model, a good fit to the CDF data is obtained if the 
proton radius is decreased~\cite{mikepc} by about a factor of $1.7$, 
as compared to the form factor model.  Now the observation has been 
extended to photoproduction as well~\cite{jetweb}.
\section{Conclusion}
In conclusion, we have shown that standard QCD
processes such as hard parton parton scattering and
soft gluon emission from valence quarks can
account for  two salient features of the total proton-proton 
cross-section, the rise at high energy and the very gentle 
decrease at low energy. 
%The main 
An important characteristic of this treatment is
that, as the minijet cross-section rises with energy,  soft gluon 
emission produces an acollinearity of the partons and reduces 
the probability of collisions. This affects the cross-sections 
in two ways : at low energy it produces a very soft decrease in 
$\sigma^{pp}_{tot}$ and contributes to the faster decrease in 
$\sigma^{p{\bar p}}_{tot}$,  at high energy it
tames the rise due to $\sigma_{jet}$. It is then possible to
have a very small $p_{tmin}$ to see the onset
of the rise around $10\div 20 \ GeV$, without encountering
too large a cross-section
when the energy climbs into the TeV range and beyond.  We stress
 that the above 
behaviour is obtained from leading soft gluon emission
from the valence quarks. Subleading emission from internal 
gluon legs is not considered here.
It is to be emphasized that singular
$\alpha_s$ appears necessary for this purpose thereby implying
that confinement plays a crucial role in the energy dependence 
of the total cross section.

Further input is needed to understand
the scale or the normalization, 
%and the Regge intercept
 which plays 
%the dominant behaviour 
a dominant role in the early decrease of the proton-antiproton cross-section.

\begin{acknowledgments}
This work was supported in part through EU RTN Contract CT2002-0311.
RG wishes to acknowledge the partial support of the the Department of 
Science and Technology, India, under project number SP/S2/K-01/2000-II.
AG acknowledges support from MCYT under project number FPA2003-09298-c02-01.
GP is thankful for the hospitality of Boston University Theoretical
Physics group, where this work was completed.
\end{acknowledgments}


\begin{thebibliography}{99}
\bibitem{mikepc}
Note that there is a misprint in \protect\cite{Borozan:2002fk} 
which states that the proton radius has to be increased by a factor 
of $1.7$; M. H. Seymour, Private Communication.
\end{thebibliography}
\end{document}